# Modeling extra-deep electromagnetic logs using a deep neural network

Sergey Alyaev[1], Mostafa Shahriari[2], David Pardo[3], Ángel Javier Omella[4], David Selvåg Larsen[5], Nazanin Jahani[1], and Erich Suter[1]

## ABSTRACT

Modern geosteering is heavily dependent on real-time interpretation of deep electromagnetic (EM) measurements. We have developed a methodology to construct a deep neural network (DNN) model trained to reproduce a full set of extra-deep EM logs consisting of 22 measurements per logging position. The model is trained in a 1D layered environment consisting of up to seven layers with different resistivity values. A commercial simulator provided by a tool vendor is used to generate a training data set. The data set size is limited because the simulator provided by the vendor is optimized for sequential execution. Therefore, we design a training data set that embraces the geologic rules and geosteering specifics supported by the forward model. We use this data set to produce an EM simulator based on a DNN without access to the proprietary information about the EM tool configuration or the original simulator source code. Despite using a relatively small training set size, the resulting DNN forward model is quite accurate for the considered examples: a multilayer synthetic case and a section of a published historical operation from the Goliat field. The observed average evaluation time of 0.15 ms per logging position makes it also suitable for future use as part of evaluation-hungry statistical and/or Monte Carlo inversion algorithms within geosteering workflows.

## INTRODUCTION

Recovering sparse hydrocarbon resources requires precise positioning of a well in a subsurface environment burdened by uncertainties. The process of dynamically changing the angle of a well based on measurements acquired and transmitted in real time (RT) from logging while drilling tools is commonly called geosteering. The most useful and used measurements for geosteering are electromagnetic (EM) measurements, which combine reliability with deep sensitivity (depth of investigation). The current generation of extra-deep EM tools (also referred to as ultra-deep EM) can detect remote boundaries up to 60 m away from measurement location (Wu et al., 2019), and recent publications have also revealed sensitivity ahead of bit (Larsen et al., 2018).

Taking advantage of such logging instruments requires automated workflows. Traditionally for EM logging, deterministic inversions are used (Sviridov et al., 2014; Pardo et al., 2015). Such processes require the solution of possibly thousands of forward problems under strict time constraints (Pardo et al., 2015; Shahriari et al., 2018). However, deterministic inversion suffers from nonuniqueness of the solutions (Shahriari et al., 2020c).

To reach high-quality geosteering decisions, it is convenient to use statistical (Bayesian) interpretation, which provides uncertainties (Kullawan et al., 2014). Bayesian workflows for updating geomodels while drilling provide promising results in synthetic testing (Chen et al., 2015; Luo et al., 2015; Alyaev et al., 2019). At the same time, adequate uncertainty representation may require up to 100,000 forward simulations per logging position (Dupuis and Denichou, 2015). In contrast to simulation software developed in the academic environment (Davydycheva et al., 2014), many existing commercial forward models have limited parallel execution capabilities. Moreover, due to the proprietary nature of the codes, they cannot be easily reimplemented for parallel execution. Thus, direct access to service companies' forward models remains a crucial element for field implementations of statistical interpretation workflows (Hermanrud et al., 2019), limiting access to Bayesian interpretation to the largest tool and service providers. It is desirable to make Bayesian interpretation accessible also to (1) operators, for geologic interpretation based on their unique knowledge of the geology of the field, which are of utmost importance for reducing uncertainties, (2) academic experts for accelerating the development of data-driven interpretation methods, and (3) smaller service companies for the development of new interpretation and geosteering workflows.

Machine learning (ML) can approximate the physical simulators based on data alone and are faster than most high-fidelity models. Deep neural networks (DNN) (He et al., 2016; Higham and Higham, 2018) have during the past few decades proven their unique ability to overcome essential challenges in various fields of science and technology (Bhanu and Kumar, 2017; Lu et al., 2017; Yu and Deng, 2017). Analogously, the research of DNN and ML techniques have also increased tremendously in the past few years in petroleum engineering and computational geophysics (Aulia et al., 2014; Hegde et al., 2015; Bougher, 2016; Lary et al., 2016; Araya-Polo et al., 2017; Bize-Forest et al., 2018; Ge et al., 2019; Puzyrev, 2019; Chen and Zhang, 2020; Colombo et al., 2020; Moghadas, 2020; Shahriari et al., 2020b; Zhu et al., 2020). Recently, there have been several noteworthy works aimed at well-log approximation: He and Misra (2019) use a neural network to predict missing dielectric dispersion logs from available logs. Shahriari et al. (2020a) and Kushnir et al. (2018) use a DNN to approximate the forward problem for a 1D medium. Because the forward problem is continuous and has a unique solution, the approximations deliver acceptable accuracy but have certain limitations. The model in Shahriari et al. (2020a) is restricted to only three layers. Kushnir et al. (2018) approximate a more complex 2D model with a geologic fault but only consider traditional deep azimuth measurements (above 400 kHz) acquired with shorter (up to 1.0 m) spacing instruments.

We present the full set of extra-deep RT EM logs consisting of 22 measurements per logging position with a DNN. To take full advantage of the depth of investigation, we train the model to respond to up to seven layers (three above and three below the logging instrument). For the training (offline) phase, we generate a data set using a commercial simulator software (Sviridov et al., 2014) that offers no access to the source code. Critical to the success of the proposed methodology is the design of a relatively small application-specific and geologically consistent training data set, for which we demonstrate that it yields accurate approximation in the modeled geosteering scenarios. For simplicity, we restrict our focus to isotropic resistivities and a logging instrument direction that is near parallel to the layering. However, a generalization to different directions and anisotropic formations is straightforward.

In this work, we create a scalable EM simulator that is suitable for cluster deployment without having access to the source code of the software or the EM tool configuration. One crucial advantage of the proposed data-driven simulator is its capability of performing more than 5000 forward evaluations per second during the execution (online) phase. In addition, the resulting DNN software overcomes limitations posed by certain commercial simulators that use graphical user interfaces, which prevent their massive parallel execution when using Markov chain Monte Carlo inversion methods.

The paper is organized as follows. We first discuss the forward model and data generation procedures for realistic EM logs. We then describe our selected DNN architecture. Finally, we present our numerical results and draw our conclusions.

## FORWARD MODEL AND TRAINING SET GENERATION

We approximate the responses of the extra-deep EM tool using a DNN. To establish the DNN model, we need a training data set. In our case, we generate the data set using a high-fidelity physics-based simulator with software provided by the tool vendor. Sviridov et al. (2014) describe the vendor's forward EM simulator, which reproduces log responses in a layered medium with arbitrary orientation of layers relative to the tool direction (the well angle). Here, we treat the forward simulator as an unknown function $\mathcal{F}$. In the following, we describe its inputs and outputs.

### Model inputs

In our study, we use the inputs native to the simulator with some assumptions to reduce the input space for the sake of training efficiency. We encode a total of seven layers (three above and three below the logging position layer), which is a number often used for practical purposes that is related to the look-around capabilities of the logging instrument. For each layer, the input contains the location of the boundary between layers relative to the measurement point (six variables — the top and bottom layers are considered to be of infinite thickness) as well as the resistivities parallel and perpendicular to the layer (7 × 2 variables). In addition, we encode the local geometry of the well by two angles. The first angle is the relative angle between the tool and the downward-pointing normal of the layered model. The second angle is the relative angle of the well 20 m ahead of the receiver. The interpolation between the angles defines the appropriate positions of the tool's transmitters required for the computation. In total, this parameterization of the model gives 22 input variables that we denote as vector $\mathbf{P} \in \mathbb{R}^{22}$.

### Synthetic logs/model outputs

The model outputs/the synthetic logs are in the same format as the logging instruments during a drilling operation. Figure 1 shows schematic of the instruments. Depending on the tool configuration, there can be up to 22 individual measurements used for RT inversion: four shallow apparent resistivities and nine pairs of deep directional measurements. We train the DNN to reproduce all of them. Assuming no azimuth (sideways) angle, we can replace the value-angle pairs of the directional measurements by projecting the directional measurements to the vertical axis, yielding one signed number. This gives a total of 13 values in the output, which we denote by $\mathbf{M} \in \mathbb{R}^{13}$. Table 1 summarizes the mnemonics used in the paper.

The RT logs are grouped in three categories:

1) The group of logs denoted as RT compensated resistivity (CRES) represent the traditional resistivity logs included in the deep EM tool (see Figure 1a). The measurements are pro-duced from two transmitter-receiver groups T1 → R2, R1 and T2 → R1, R2 with the symmetric resistivity compensation (Larsen et al., 2019). The phase difference and attenuation of the signals are recorded at 2 MHz and 400 kHz (see Table 1). These nonazimuthal measurements have a sensitivity up to 5.0 m radial distance from the logging instrument (Meyer et al., 2008). The measurements are transformed into apparent resistivities (in ohm·m) using corrections described by Meyer (1999).

2) The group denoted as RT ARSLLM is an azimuthal measurement produced from the deep resistivity logging instrument between two transmitter-receiver pairs T2 → R3 and T1 → R4 operating at 400 kHz (see Figure 1a). The sampled voltages are processed internally in the tool using standard symmetric compensation and provide signal strength and a target direction, which implies the excess of conductivity in the EM environment around the transmitter-receiver coil (Meyer et al., 2008; Fang et al., 2010). In a 1D environment, the maximum signal is always oriented in the direction of the layering. Thus, we omit the angle and use the signed value of the signal. The processing provides real and imaginary components, which are presented in nanovolts (nV). Here, we only model the imaginary component, which is more useful for geosteering, and it is provided in RT.

3) The extra deep EM tool (see Figure 1b) operates at frequencies of 20 and 50 kHz and deploys a set of synchronized receivers and two transmitters, one which is cross-oriented for acquiring the azimuthal components (Hartmann et al., 2014; Larsen et al., 2015). Its inputs are denoted as RT EDAR

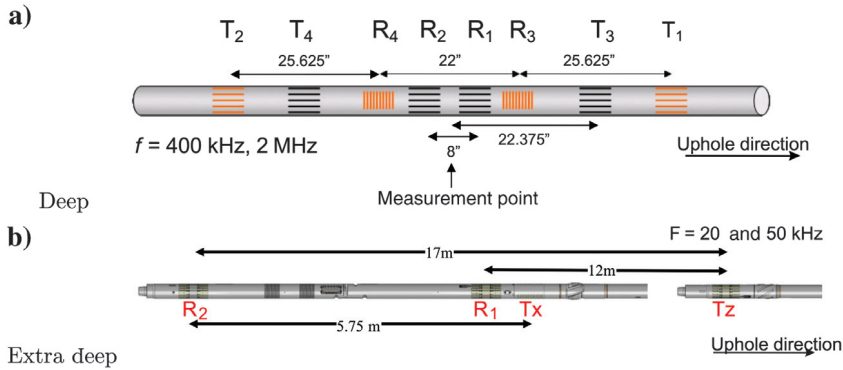

Figure 1. The schematic of the EM downhole tools modeled in this paper. The "Deep" EM tool operates at 2 MHz and 400 kHz, whereas the "Extra Deep" EM tool operates at 20 and 50 kHz. The transmitters and receivers are denoted with T and R, respectively. The horizontal lines indicate coaxial coils, whereas the vertical lines indicate transverse-oriented coils.

Table 1. Description of mnemonics of the signals used in the paper.

| | | | Mnemonics' meanings | | |
|---|---|---|---|---|---|
| Mnemonic | Tool | T → R | Description | Units | Frequency |
| RT_CRES_RPCEHX | Deep | T1 → R1, R2, T2 → R1, R2 | AR: phase difference, cor/com | ohm m | 2 MHz |
| RT_CRES_RACEHX | Deep | T1 → R1, R2, T2 → R1, R2 | AR: attenuation, cor/com | ohm m | 2 MHz |
| RT_CRES_RPCELX | Deep | T1 → R1, R2, T2 → R1, R2 | AR: phase difference, cor/com | ohm m | 400 kHz |
| RT_CRES_RACELX | Deep | T1 → R1, R2, T2 → R1, R2 | AR: attenuation, cor/com | ohm m | 400 kHz |
| RT_ARSLLM | Deep | T2 → R3, T1 → R4 | AMS: imaginary, com | *nV | 400 kHz |
| RT_DTK_ATC50X | Extra Deep | Tz → R1, R2 | PR: attenuation, com | *dB | 50 kHz |
| RT_DTK_PDC50X | Extra Deep | Tz → R1, R2 | PR: phase difference, com | *degree | 50 kHz |
| RT_EDAR_ImV50k | Extra Deep | Tx → R2 | AMS: imaginary, com | *nV | 50 kHz |
| RT_EDAR_ReV50k | Extra Deep | Tx → R2 | AMS: real | *nV | 50 kHz |
| RT_DTK_ATC20X | Extra Deep | Tz → R1, R2 | PR: attenuation, com | *dB | 20 kHz |
| RT_DTK_PDC20X | Extra Deep | Tz → R1, R2 | PR: phase difference, com | *degree | 20 kHz |
| RT_EDAR_ImV20k | Extra Deep | Tx → R2 | AMS: imaginary | *nV | 20 kHz |
| RT_EDAR_ReV20k | Extra Deep | Tx → R2 | AMS: real | *nV | 20 kHz |

The corresponding schematics of the Deep and Extra Deep tools can be found in Figure 1. Here, measurements marked with a star are scaled to nondimensional units and the ± sign is used to distinguish the up or down direction of the maximum signal for AMS measurements.
AR, apparent resistivity (nonazimuthal); AMS, azimuthal maximum signal; PR, propagation resistivity (nonazimuthal); "cor", borehole correction and correction from the complex refractive index method; "com", compensation using extra recievers; RT, transmitted in real time; Ti, transmitter; "i": its id on the tool schematic; and Ri, reciever; "i": its id on the tool schematic.

and RT DTK for the azimuthal (Tx to R2) and nonazimuthal (Tz to R1 and R2) components, respectively. For EDAR, the signals are provided in nV and we use the ± sign to distinguish the maximum signal direction. For DTK, the data from the two receivers of the nonazimuthal signal are used to compensate for any receiver sensitivity variation using special design and proprietary technologies (Larsen et al., 2019); the raw signal in decibel (dB) is used.

### Ground truth data set

The ground truth data set for this study is generated by an automatic workflow. For each input, we first select a random realization of inputs following the rules described below. Then, an AutoIt script (Bennett, 1999) executes the proprietary modeling tool for the selected input models, which produces noise-free logs with the same preprocessing that is used in the logging tool.

For this study, we restrict our focus to the angle range that is most important for geosteering in near-horizontal layering: 80°–92°, where 90° means drilling parallel to the layers. We assume that the well trajectory is straight around the measurement point, meaning that no bending in the logging instrument occurs and the transmitters have the same orientation as the receivers. Thus, the two angle parameters are equal. All inner layers in all models have thicknesses uniformly distributed between 0.3 and 20.0 m and rounded to 0.1 m. In addition, for simplicity, we restrict our study to isotropic resistivities.

The initial training data set containing 22,469 samples consists of a generic layered medium with uniformly distributed resistivities in the logarithmic scale, yielding values between 1 and 122 ohm·m for each of the layers (see Figure 2a).

Then, we enrich the training data set with 50,492 additional samples containing alternating sand-shale layers that better capture the variety in high contrasts rather than small heterogeneities across the layers. The ability to produce accurate predictions in such high-contrast scenarios is essential and of great value during geosteering operations, so the ML model must be trained with a large number of samples to improve its quality for this specific geologic scenario. For that, we consider different sand-shale sequences typical to, e.g., fluvial reservoirs (Larsen et al., 2015), where the shales resistivities are in the range of 0.9–4.1 ohm·m and the sand resistivities are in the range of 49.4–221.4 ohm·m. Within these ranges, resistivities are randomly sampled in the logarithmic scale. We consider two cases with respect to the nature of the middle (logging position) layer. In the case in which the logging position is in a sand (see Figure 2b) we generate 25,203 samples with seven alternating layers. In the case in which the logging position is in a shale (see Figure 2c) we generate 25,289 samples with five alternating sand-shale layers. Because the sensitivity to the outermost (first and seventh) layers is very low, the model is simplified to only five layers in this scenario. We implement this by imposing that the two top layers share their electrical properties. The same is imposed on the two bottom layers (see Figure 2c). The total of 72,961 samples described above comprise what we call the *basic data set* $\mathbf{D}^b$.

Early numerical experiments revealed that the data set dominated by alternating sand-shale layers lacks the data needed to accurately approximate the measurements in thick shale, which is typical when approaching a reservoir. Therefore, the data set is augmented with 11,709 samples of "semidegenerate" sand-shale layering. This part of the data set is generated with the same rules as the high-contrast layers, but for each sample a random number of layers near the top or bottom were assigned the resistivity of the first and seventh layer, respectively. The logging position layer always maintains an independent resistivity. The final *extended data set*, denoted $\mathbf{D}^e$, contains 84,599 input-output pairs. We denote the $i$th input-output pair (sample) as $(\mathbf{P}_i, \mathbf{M}_i) \in \mathbb{R}^{22} \times \mathbb{R}^{13}$.

The extremely high values of the apparent resistivities near the layer boundaries (often called "horns") are truncated by the simulator to 2000 ohm·m. Our experiments indicate that removing such samples from the data set provides a better fit of the final results.

The remaining 78,877 samples are split into training (80%), validation (10%), and testing (10%) data sets. Testing samples are unused during training, and they are used only for quality assessment of the model in the "Crossplots" subsection.

## DEEP-LEARNING APPROACH

In this work, we design a DNN-based approximation $\mathcal{F}_w$ (where $w$ stands for the finite set of weights) of the forward function $\mathcal{F}$. In the following, we describe the ML algorithm including its rescaling, loss function, architecture, and training process.

### Rescaling the data

To equalize the effect of each component of the training data set during the optimization process, we build a min-max linear rescaling over each variable. Note that for input and output variables representing resistivities, we take the logarithm prior to rescaling. We denote these rescaling functions and their inverses as

$$\Psi_p := \{\Psi_p^1 \ldots, \Psi_p^p\}, \quad \Psi_m := \{\Psi_m^1 \ldots, \Psi_m^m\},$$
$$\Psi_p : (\mathbb{R}^p) \to [0.5, 1.5]^p, \quad \Psi_m : (\mathbb{R}^m) \to [0.5, 1.5]^m,$$
$$\mathbf{P}'_i = \Psi_p(\mathbf{P}_i), \qquad \mathbf{M}'_i = \Psi_m(\mathbf{M}_i),$$
$$\mathbf{P}_i = \Psi_p^{-1}(\mathbf{P}'_i), \qquad \mathbf{M}_i = \Psi_m^{-1}(\mathbf{M}'_i), \qquad (1)$$

where $p$ and $m$ are the number of variables (components) in $\mathbf{P}$ and $\mathbf{M}$, respectively.

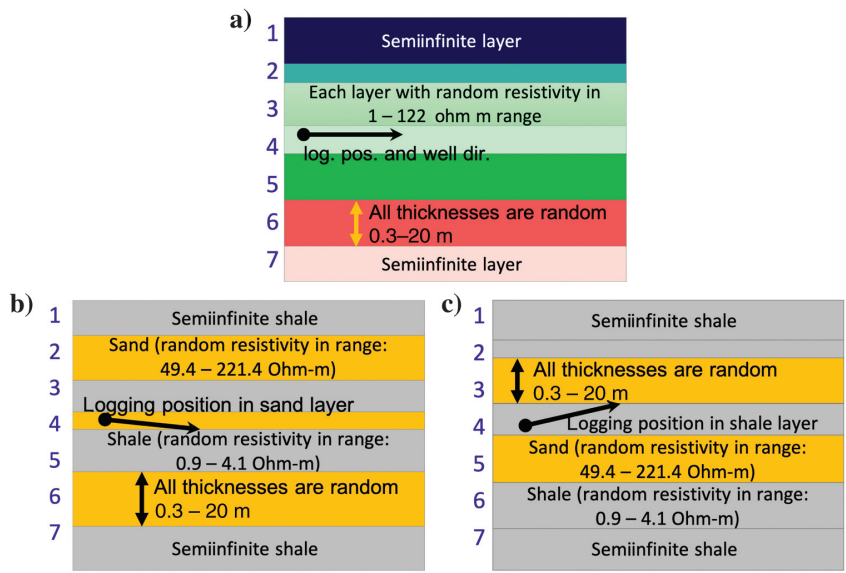

Figure 2. Geologic model schematics. The measuring position denoted by a dot with a well direction arrow is located in the middle layer. (b) Sand as the middle layer. (c) Shale as the middle layer.

The functional approximation of the forward model $\mathcal{F}_w$ consists of the rescaling introduced in equation 1 and the DNN relation between rescaled inputs and outputs $\tilde{\mathcal{F}}_w$, which can be defined as

$$\mathcal{F}_w(\mathbf{P}) = \Psi_m^{-1} \mathbf{M}' = \Psi_m^{-1}[\tilde{\mathcal{F}}_w(\mathbf{P}')], \qquad (2)$$

where $\mathbf{M}'$ is the rescaled DNN approximation of the output.

### Loss function

We want the DNN approximation to satisfy $\mathcal{F}_w(\mathbf{P}'_i) \approx \mathbf{M}'_i$ for all $(\mathbf{P}'_i, \mathbf{M}'_i) \in S'_t \cup S'_v$.
Therefore, we select the following loss function:

$$L(\tilde{\mathcal{F}}_w(\mathbf{P}'_i, \mathbf{M}'_i)) = \|\tilde{\mathcal{F}}_w(\mathbf{P}'_i) - \mathbf{M}'_i\|_1, \qquad (3)$$

where $\|\|_1$ is the $l_1$ norm.

### Architecture

We use the architecture shown in Figure 3:

$$\tilde{\mathcal{F}}_w := \mathcal{L}_{w_5} \circ \mathcal{B}_{w_4}^4 \circ \mathcal{B}_{w_3}^3 \circ \cdots \circ \mathcal{B}_{w_0}^0, \qquad (4)$$

where $\circ$ is the function composition operator defined as $f \circ g(x) \equiv f(g(x))$. In the above, $w = \{w_i, i = 0, \ldots, 5\}$ are the weights (unknowns) of the DNN. Then, the blocks of the DNN are defined as

$$\mathcal{B}_{w_i}^i := (\mathbf{n} \circ \mathbf{l}_{w_i^2}^\mathbf{c} + \mathcal{I})(\mathbf{n} \circ \mathbf{l}_{w_i^1}^\mathbf{c}) \quad \forall i \in \{0, 1, \ldots, 4\},$$
$$\mathcal{L}_{w_5} := (\mathbf{n} \circ \mathbf{l}_{w_5^2}^\mathbf{d} \circ \mathbf{R} \circ \mathbf{n} \circ \mathbf{l}_{w_5^1}^\mathbf{t}), \qquad (5)$$

where $w_{(\cdot)} = \{w_{(\cdot)}^1, w_{(\cdot)}^2\}$ are the weights, $\mathcal{I}$ is the identity operator, and the function $\mathbf{l}_{w_i}^\mathbf{c}$ is a convolutional layer with its kernel size equal to three and its filter size being $40 \times i$. A convolutional layer extracts the most salient features of its inputs, whereas it is computationally more efficient than a fully connected layer. The blocks mentioned above resemble a residual block, which enhances convergence and maximizes numerical stability.

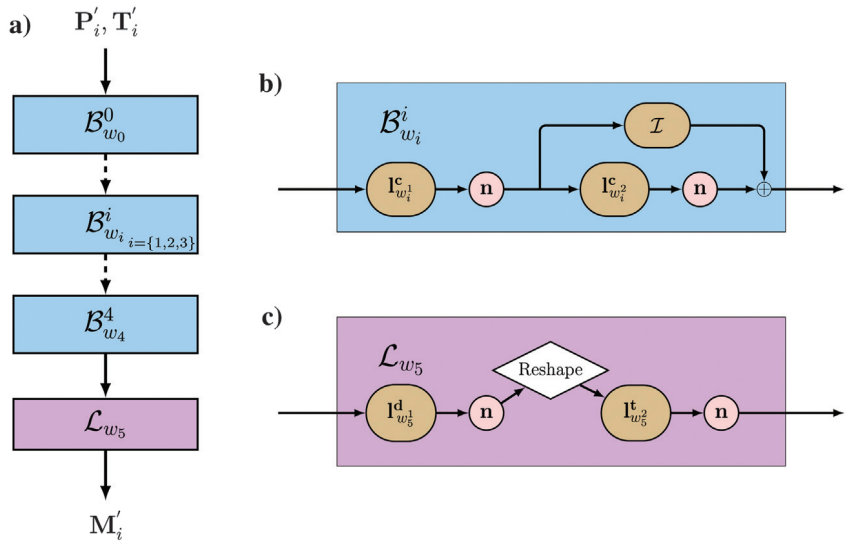

Figure 3. (a) Sketch of the DNN architecture composed by the addition of five blocks of type $\mathcal{B}$ and one block of type $\mathcal{L}$. (b) Architecture sketch of $\mathcal{B}$-type block. (c) Architecture sketch of $\mathcal{L}$-type block.

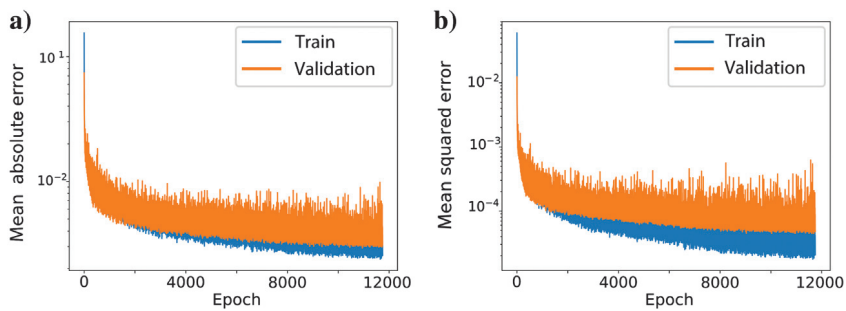

Figure 4. History of decrease of (a) mean absolute and (b) mean squared error during training.

 

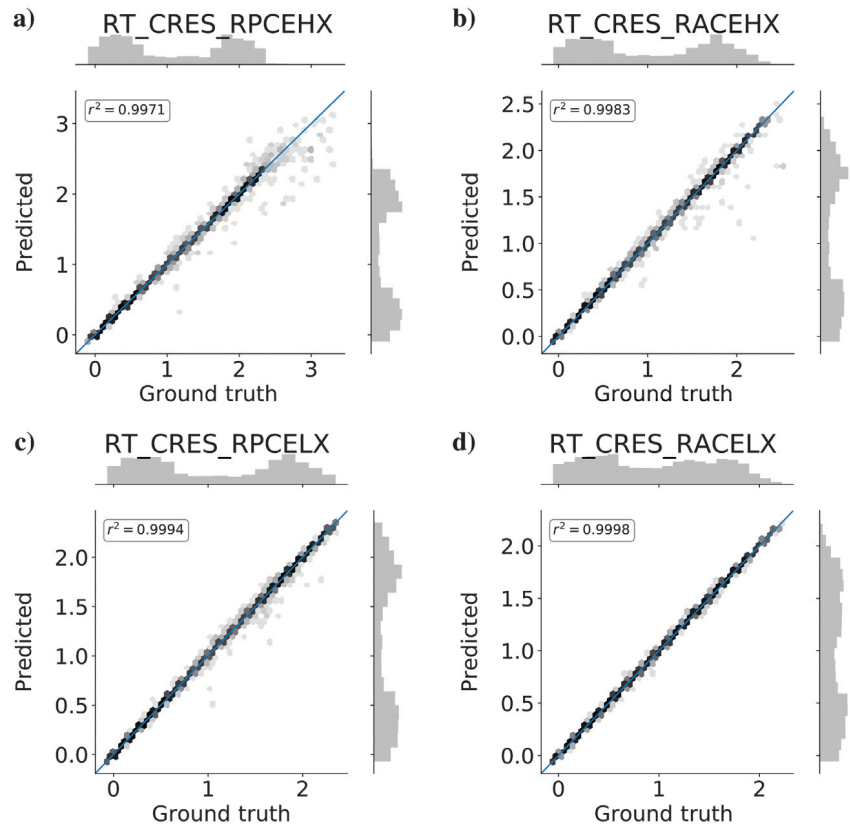

Figure 5. Crossplots of the predictions on test data versus physical simulation for shallow resistivity logs. Resistivity estimated from (a) phase difference and (b) attenuation at 2 MHz frequency and (c) phase difference and (d) attenuation at 400 kHz frequency. T1 → R2 and T2 → R1 pairs are used (see Figure 1a and Table 1). Note the logarithmic scale.

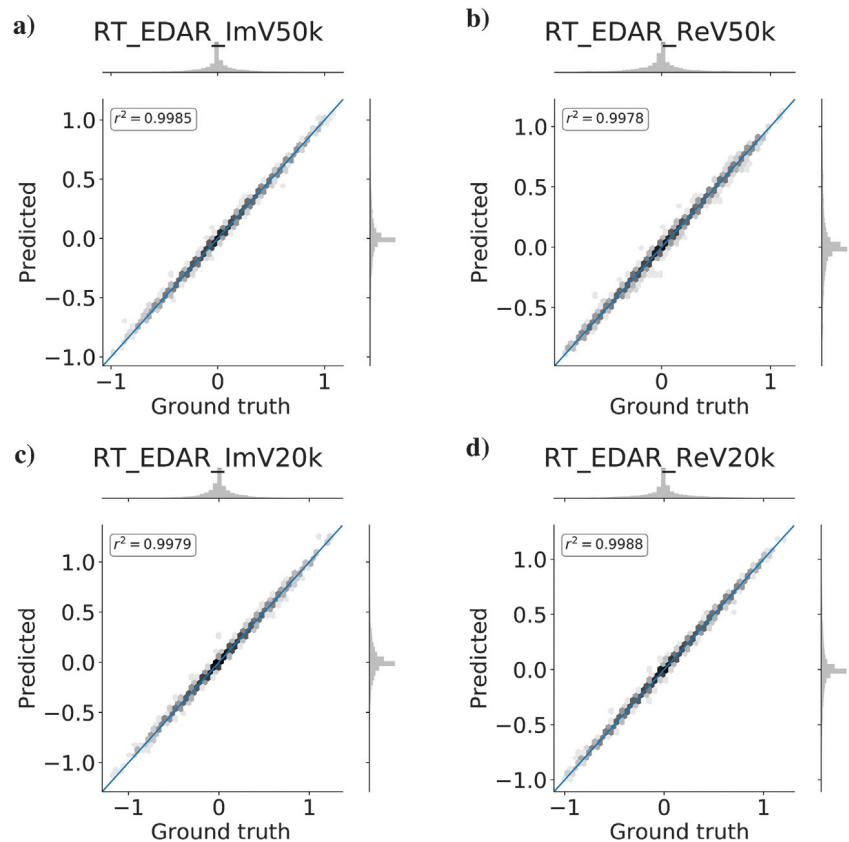

Figure 6. Crossplots of the predictions on test data versus physical simulation for the azimuthal extra-deep measurements in nV: 50 kHz (a) imaginary and (b) real and 20 kHz (c) imaginary and (d) real. Tx → R2 pair is used (see Figure 1b and Table 1). The measurements' scales have been normalized to nondimensional units.

The terms $\mathbf{l}^d$ and $\mathbf{l}^t$ are fully connected layers, which perform the final feature extraction and downsample the output to the required output dimensions. The term $\mathbf{R}$ is a function that reshapes its input behind the last layer, and $\mathbf{n}$ denotes the activation function. An activation function is a known simple nonlinear function applied on its vector input component wise, e.g., a sigmoid or a rectified linear unit (ReLU). In this work, we consider the so-called ReLU activation function:

$$\mathbf{n}(x_1,\ldots,x_r) = (\max(0,x_1),\ldots,\max(0,x_r)). \quad (6)$$

## Training

The described architecture with depth $N = 5$ gives a total of 462,453 trainable parameters, also known as weights, which are determined by minimizing the misfit on the training data set. This number of weights is sufficiently large to ensure good-quality results. Smaller architectures are probably possible (e.g., Zhang et al. [2017] advocate limiting the number of weights in classification problems, although such theoretical results are not directly applicable here). The use of a large number of weights may lead to overfitting problems. To prevent that, we evaluate our DNN over the validation data set after each epoch to assess the quality of generalization. The training is terminated when the error on the validation data set stops decreasing.

After successful training of the DNN offline, the process of computing the forward function during RT online operation reduces to evaluating a DNN. The online evaluation only requires several simple algebraic operations proportional to the number of weights, and it is computationally very fast.

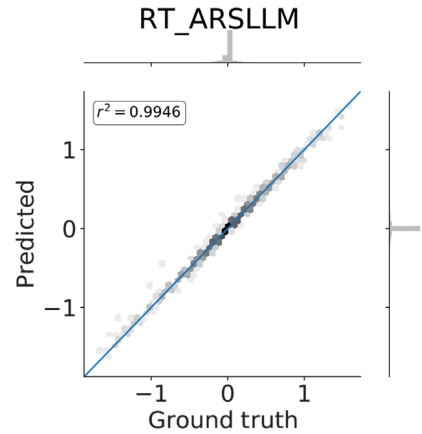

Figure 8. A cross-plot of the predictions on test-data vs physical simulation for the Deep 400 kHz azimuthal imaginary log. T1 → R4 and T2 → R3 pairs are used, see Figure 1a and Table 1.

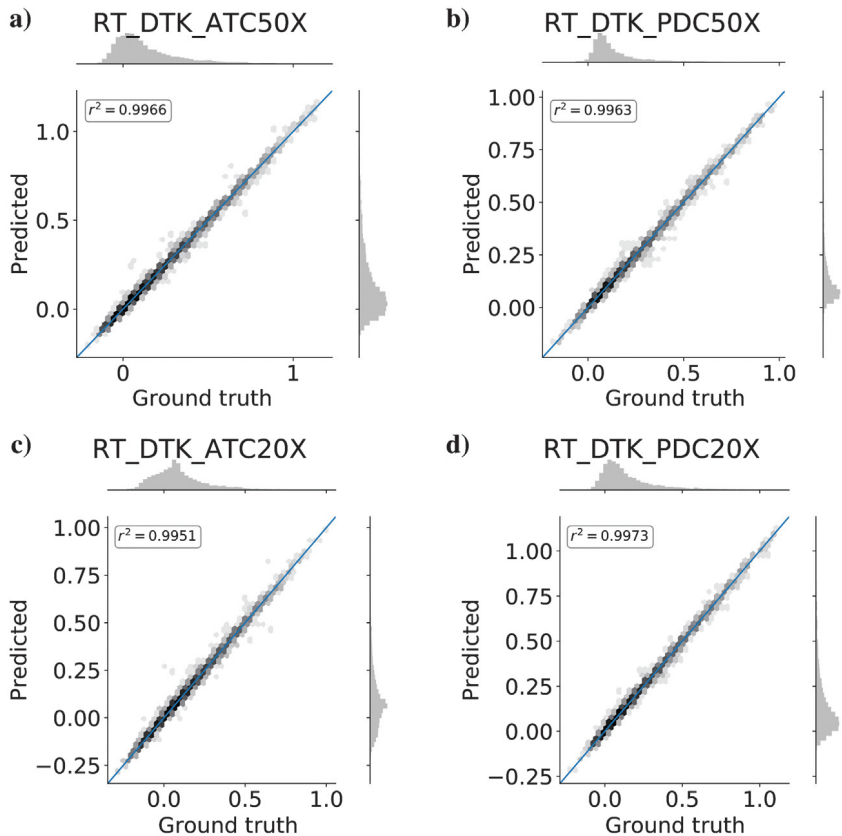

Figure 7. Crossplots of the predictions on test data versus physical simulation for the nonazimuthal extra-deep measurements: (a) attenuation, dB and (b) phase difference, degrees for 50 kHz and (c and d) for 20 kHz. Tz → R1 and Tz → R2 pairs are used (see Figure 1b and Table 1). The measurements' scales have been normalized to nondimensional units.

## NUMERICAL RESULTS

In this section, we present a numerical verification of the methodology. First, we provide the details of the used training setup. Then, we verify the fit on the testing data set, which was excluded from training data and stopping criteria. After that, we test the DNN model on an independent synthetic example inspired by real geosteering operations in layer-cake geology. In this synthetic example, we also compare the performance of another DNN model trained on a smaller data set $\mathbf{D}^b$ that does not cover all scenarios from the synthetic example. Finally, we show that the trained DNN model provides sufficiently accurate results for a realistic case from the Goliat field.

### Training setup

We train our DNN over multiple epochs on the training data set from the *extended data set* $\mathbf{D}^e$ and batch size equal to 512. We use the default settings of the Adam optimizer (Kingma and Ba, 2015) in TensorFlow (Abadi et al., 2015). Then, we use the validation data set for early stopping to prevent overfitting using the standard EarlyStopping callback in TensorFlow with the patience (the amount of epochs without decrease of the loss for the validation data set) set to 800. Figure 4 shows the reduction of training and validation errors. The early stopping was triggered after 11,745 epochs with the total training phase taking approximately 13 h on a consumer-grade GPU (RTX 2080ti). We use the weights obtained at epoch 10,946 to produce the verification results in the rest of this section.

### Crossplots

Although the training/validation data sets ensure that the model approximates the simulator reasonably well, they do not ensure that the model will generalize to other data points. Nor does the average error explain how well different measurements are reproduced.

Figures 5, 6, 7, and 8 present the data fit for all the 13 modeled RT measurements. The crossplots show deviation of the predictions from the original high-fidelity forward model. The straight 45° diagonal corresponds to the perfect fit. To give a numerical estimate of the quality of the fit, we compute the coefficient of determination $r^2$ for each of the measurements.

The evaluation on the test data gives the $r^2$ coefficient of 0.99 and above demonstrating a well converged model. One can observe that the DNN delivers a better data fit for shallow logs (Figure 5) than for deeper nonazimuthal measurements (see Figure 7). One explanation can be that these logs can be well approximated with a weakly nonlinear averaging, which a DNN can easily pick up.

### Synthetic example

To give an engineering insight into the quality of the model, in this section we generate a synthetic log for all measurements in a layer-cake environment for a near-horizontal well. The well trajectory and the horizontal layer-cake geomodel are presented in Figure 9. The log was sampled with a point each meter along the measured depth of the well, yielding a total of 901 logging positions.

This example represents a reasonably typical task for a predictive model in a RT operation. Therefore, it is important to verify the DNN's capability to provide quick and accurate results for such a setup.

The DNN is evaluated on a workstation running Intel Xeon W-2155 CPU with 10 3.30 GHz cores. For the full range of 901 logging positions along the well, the evaluation takes 0.13 s, that is, 0.15 ms per logging position, which should suffice even for the most demanding RT inversion algorithms. The performance results are similar to those reported in Kushnir et al. (2018).

In Figures 10, 11, 12, and 13, we compare the logs obtained from the original physical simulation versus those predicted by the trained DNN. All plots qualitatively coincide except for some small discrepancies appearing near the layer interfaces.

Because the example is generated separately from the training data set, it potentially contains combinations of values of angles, resistivities, and depths that may not be present in the training set.

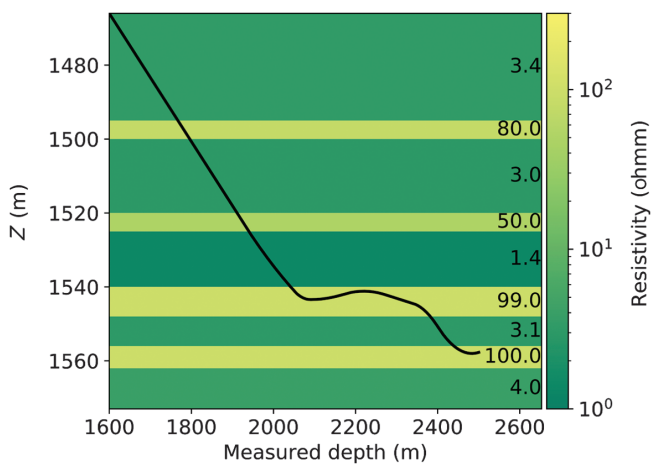

Figure 9. The well trajectory and layer resistivities for the realistic log example in measured depth versus true vertical depth coordinates.

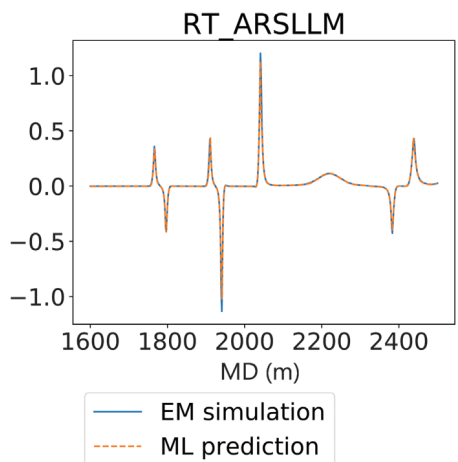

Figure 10. A comparison of ML-based vs physical approximations of this log for the realistic test case in Figure 9. T1 → R4 and T2 → R3 pairs are used, see Figure 1a and Table 1. The measurements' scale has been normalized.

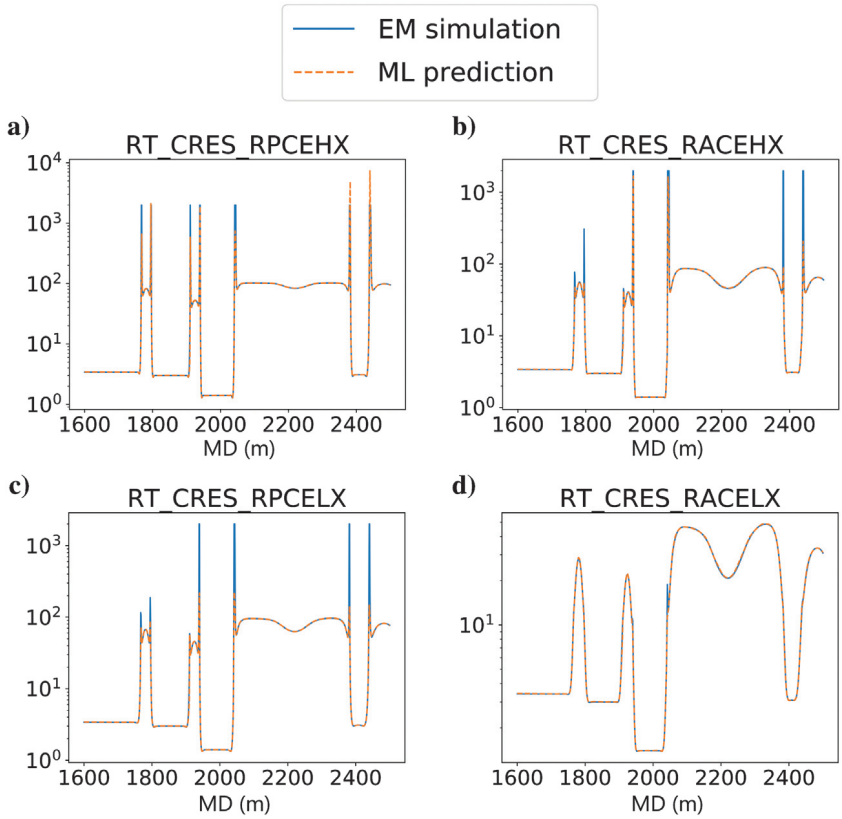

Figure 11. A comparison of ML-based versus physical approximations for shallow resistivity logs on the realistic test from Figure 9. Resistivity estimated from (a) phase difference and (b) attenuation at 2 MHz frequency and (c) phase difference and (d) attenuation at 400 kHz frequency. Note the logarithmic scale.

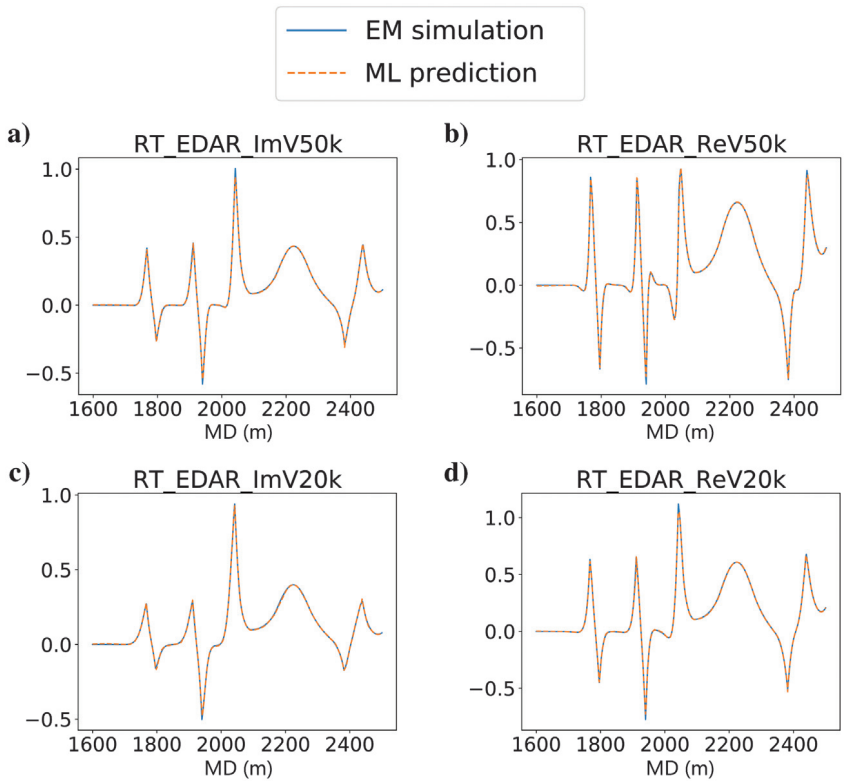

Figure 12. A comparison of ML-based versus physical approximations for the azimuthal extra-deep measurements in nV: 50 kHz (a) imaginary and (b) real and 20 kHz (c) imaginary and (d) real. Tx → R2 pair is used (see Figure 1b and Table 1). The measurements' scales have been normalized to nondimensional units.

Specifically, we observe pronounced inconsistencies before the landing in the reservoir for the deeper nonazimuthal logs (see Figure 13). This might have several explanations. First, there may be an ambiguity of layer thicknesses when a degenerate model (containing fewer layers) is converted to a seven-layer model during training and evaluation. Consequently, the measurements that have learned sensitivity to more layers are affected the most. Second, the training data contain a relatively low number of samples in thick shale above a reservoir. To that end, it is important to ensure that the model improves with the availability of relevant training data that better span the possible configurations.

In addition, Figure 13 shows the synthetic logs created by the same DNN architecture, which was trained on the smaller *basic data set* $\mathbf{D}^b$, which completely omits the semi-degenerate samples that mimic the drilling in thick shales above reservoir. The results in the initial 150 m measured depth (MD) in these attenuation logs are clearly misrepresented by the DNN trained on the smaller data set. We also observe a mismatch on other deep measurements, but it is less pronounced compared with the attenuation logs. This improvement indicates that the proposed architecture is suited for practical logging scenarios, and that the model quality improves with additional training data relevant for examples at hand. At the same time, increasing the training data set might be insufficient. We emphasize that to get high-quality data-driven approximations for this example, we designed a data set that captured the logging in thick shale layers.

## Goliat field example

To further verify that the resulting DNN can generalize, in this section we consider an example from the Goliat field in the Barents Sea. This fluvial formation of Middle Triassic age is characterized by a prograding delta front environment (Larsen et al., 2015). We select a section of the earth model that has no significant 2D features between MD x000 and x105 (see Figure 8 in Larsen et al., 2015). Therefore, we approximate it by a 1D model (see Figure 14). As before, in this study, we do not consider anisotropies.

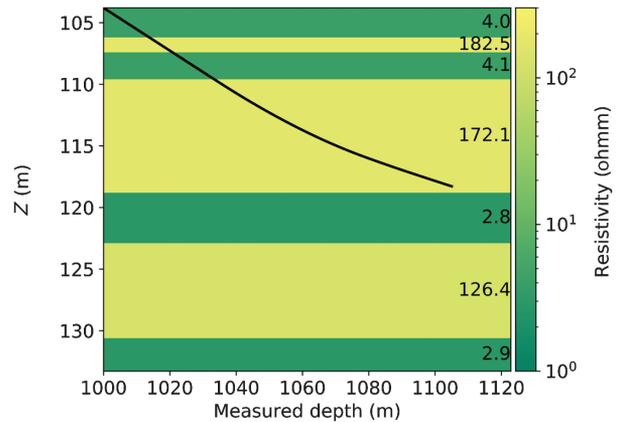

Figure 14. A layer-cake approximation of the geology near the well-landing point for well A from Larsen et al. (2015). The earth model and the well trajectory have been rotated five degrees to make the layering horizontal; here, MD 1000 corresponds to x000 in the published inverted model.

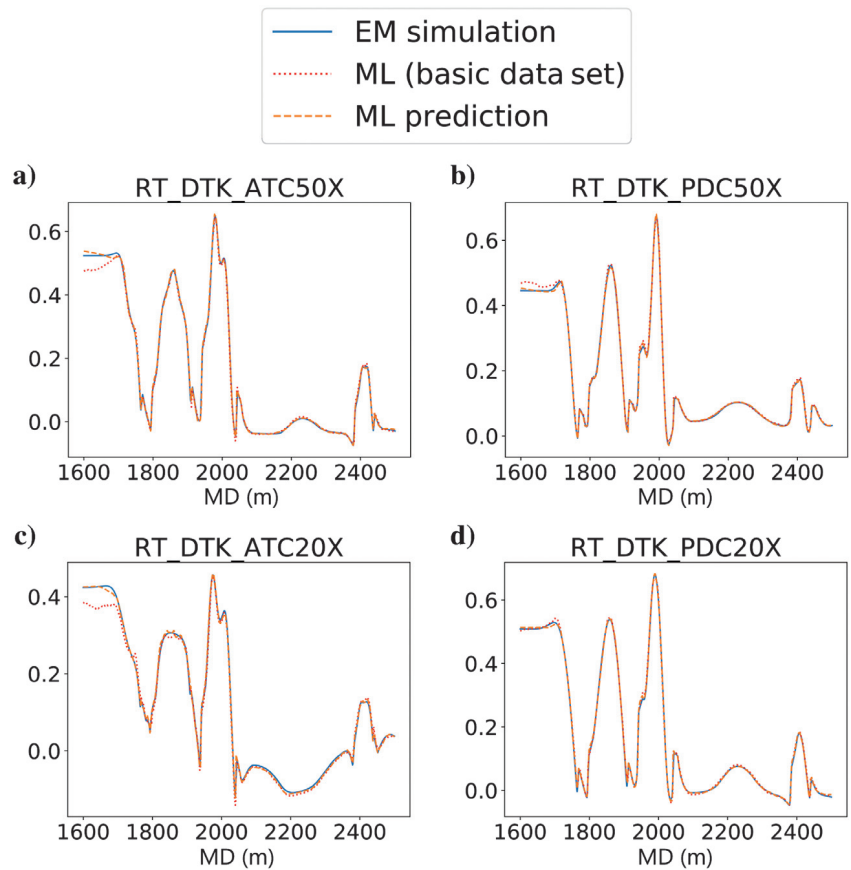

Figure 13. A comparison of ML-based versus physical approximations for the nonazimuthal extra-deep measurements: (a) attenuation, dB and (b) phase difference, degrees for 50 kHz and (c and d) for 20 kHz. $Tz \to R1$ and $Tz \to R2$ pairs are used (see Figure 1b and Table 1). The measurements' scales have been normalized to nondimensional units. The figures also show a DNN approximation based on the basic data set.

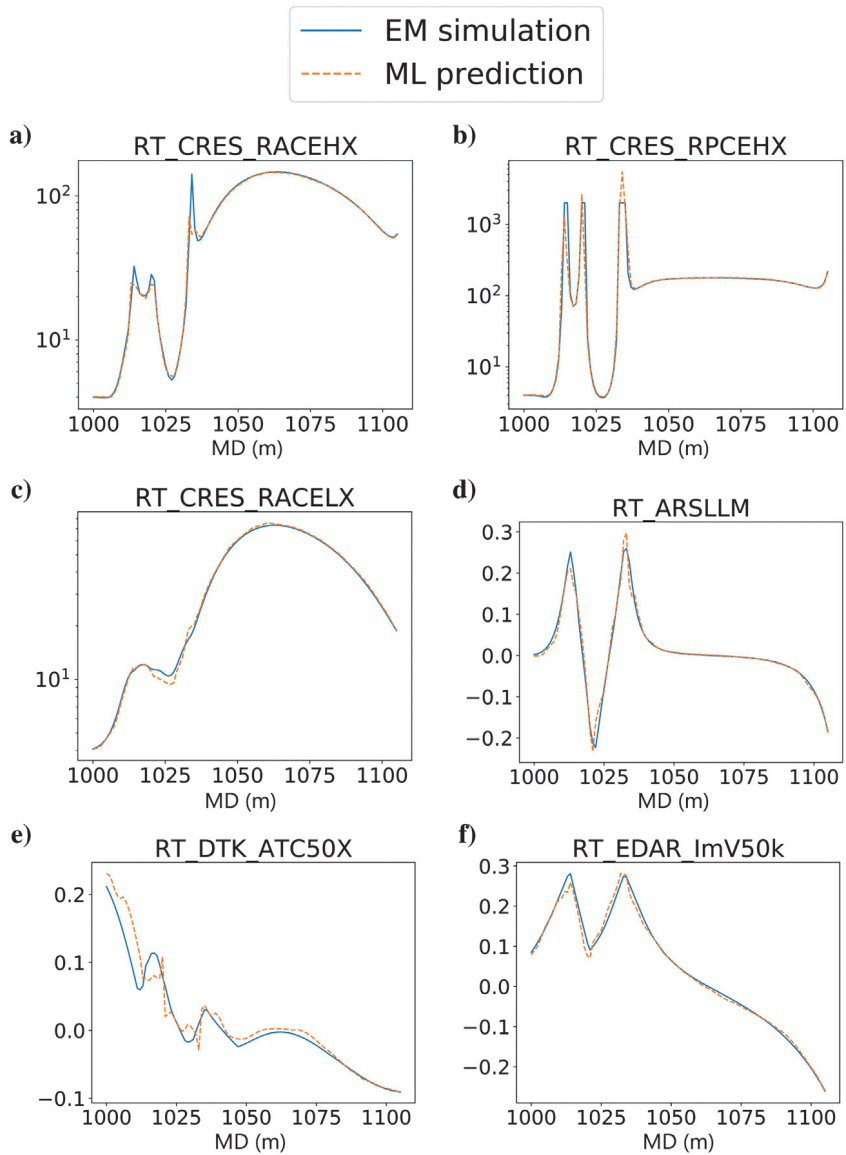

Figure 15. Comparison of six logs using a commercial EM simulator versus its ML approximation when applied to the section of the "Goliat Field example" described in Figure 14. The mnemonics are summarized in Table 1.

Figure 15 compares the simulation of six representative logs using the original commercial simulator and its DNN approximation for the considered Goliat field example. The results for the synthetic layered case from the previous example (see Figures 10, 11, 12a, 13a) are of superior quality compared with those in Figure 15. However, for the synthetic case, we have extended the original *basic data set* with samples covering the problematic areas. This example highlights that the results achieved with no tuning of DNN or training data set still exhibit acceptable accuracy also for field cases. The accuracy is specifically good for shallow logs that exhibit smaller depths of investigation. As in the previous example, we expect that further tailoring the data set to the geology at hand should improve the accuracy.

## CONCLUSION

In this work, we have demonstrated that a DNN can provide a high-quality approximation to a complex, industry-quality forward model for extra-deep EM logs used in modern geosteering operations. The DNN was trained using data generated with a numerical simulator provided by the logging instrument vendor but without access to or knowledge of its source code or the configuration of proprietary tools.

We considered a relatively small data set composed of 63,122 samples to train a high-dimensional function $\mathcal{F}_w: \mathbb{R}^{22} \to \mathbb{R}^{13}$. By making the data set geologically consistent and application oriented, we produced a good approximation to the relevant logs acquired during a simulated geosteering operation on a synthetic case and historical well from the Goliat Field. At the same time, our numerical examples show that the approximation works best within the regions that are well represented by the training data set. Therefore, an extension of the data set with expected geologic configurations is recommended to maximize performance in each specific geosteering operation. Future extensions include considering lower angles typical for well landing, anisotropic resistivities, and more complex geologic configurations.

We assert that data-driven approximations as the one presented in this work are of crucial importance when it comes to future implementations of advanced inversion and interpretation workflows. The workflows based on a DNN no longer require extra implementation on the side of the instrument vendor and hence become accessible to wider range of independent companies and researchers. In addition, the resulting DNN forward model can be rapidly evaluated (in our work, it takes 0.15 ms to evaluate all logs per logging position on a regular workstation). Such short execution times open the door for potential use as part of evaluation-hungry statistical and/or Monte Carlo inversion algorithms.

Future work toward the inversion will include a study of the DNN performance with noisy data as well as tailoring of bias correction algorithms for the ML. Such studies will help to distinguish unhandled geologic complexity from biases and inaccuracies generated by the ML algorithms.

## ACKNOWLEDGMENTS

The summaries of the tools are adapted from SPWLA's Resistivity Special Interest Group (Rt-SIG), see web page: https://

www.spwla.org/SPWLA/Chapters_SIGs/SIGs/Resistivity_/Resistivity.aspx (accessed on 10.09.2020).

Researchers at NORCE are supported by the research project "Geosteering for IOR" (NFR-Petromaks2 project no. 268122), which is funded by the Research Council of Norway, Aker BP, Equinor, Vår Energi, and Baker Hughes Norway.

M. Shahriari has been funded by the Austrian Ministry for Transport, Innovation and Technology (BMVIT), the Federal Ministry for Digital and Economic Affairs (BMDW), and the Province of Upper Austria in the frame of the COMET-Competence Centers for Excellent Technologies Program managed by Austrian Research Promotion Agency FFG and the COMET Module S3AI.

D. Pardo has received funding from the European Union's Horizon 2020 research and innovation programme under the Marie Sklodowska-Curie grant agreement no. 777778 (MATHROCKS), the European POCTEFA 2014-2020 Project PIXIL (EFA362/19) by the European Regional Development Fund (ERDF) through the Interreg V-A Spain-France-Andorra programme, the Project of the Spanish Ministry of Economy and Competitiveness with reference MTM2016-76329-R (AEI/FEDER, EU), the BCAM "Severo Ochoa" accreditation of excellence (SEV-2017-0718), and the Basque Government through the BERC 2018-2021 program, the two Elkartek projects 3KIA (KK-2020/00049) and MATHEO (KK-2019-00085), the grant "Artificial Intelligence in BCAM number EXP. 2019/00432," and the Consolidated Research Group MATHMODE (IT1294-19) given by the Department of Education.

## DATA AND MATERIALS AVAILABILITY

This work is based on synthetic data. The data set cannot be fully disclosed due to the proprietary nature of the modeled tool.

Biographies and photographs of the authors are not available.